\documentclass[12pt]{iopart}

\usepackage{amsmath}
\usepackage{iopams}
\usepackage{multicol,float,graphicx,color,epstopdf}

\begin{document}

\maketitle

\title{Spectra of Reduced Fractals and Their Applications in Biology}

\author{D. T. Pham}
\address{Department of Biology, The University of Texas at Arlington, TX 76019, USA}
\ead{npham@mavs.uta.edu}

\author{Z. E. Musielak}
\address{Department of Physics, The University of Texas at Arlington, TX 76019, USA}
\ead{zmusielak@uta.edu}

\begin{abstract}
Fractals with different levels of self-similarity and magnification are defined 
as reduced fractals.  It is shown that spectra of these reduced fractals can be constructed 
and used to describe levels of complexity of natural phenomena. Specific applications 
to biological systems, such as green algae, are performed, and it is suggested that the 
obtained spectra can be used to classify the considered algae by identifying spectra 
associated with them.  The ranges of these spectra for green algae are determined 
and their extension to other biological as well as other natural systems is~proposed.
\end{abstract}

\section{Introduction}\label{sec1}

Mandelbrot~\cite{1} originally defined fractals as objects that are self-similar 
on all scales and whose dimensions are different than their topological 
dimensions.  From a mathematical perspective, fractals are defined by 
listing their four basic characteristics~\cite{1,2,3}, which are: self-similarity 
at all scales, fine structure at all magnifications, too irregular to be 
described by Euclidean geometry, and have non-topological (Hausdorff) 
dimension; mathematical objects that have these characteristics are 
called classical fractals ~\cite{4}. To make fractals applicable to 
Nature, Mandelbrot~\cite{1} changed the definition to a more casual: 
“A fractal is a shape made of parts similar to the whole in some way”.  
Thus, the main difference between classical fractals and fractals that 
obey Mandelbrot's casual definition is that their self-similarity is exact 
for the former and non-exact for the latter~\cite{3,4}.

Fractals have been used in numerous research topics ranging from biology 
and biomedicine to physics, astronomy, geology, computer science, and in 
epidemiology, emerging diseases as well as in comparative studies~\cite{1,2,3,4,5,6}.
In the previous applications of fractals to biology~\cite{7,8,9,10,11}, the main emphasis 
was given to population biology~\cite{12}, plant structures~\cite{13}, gene expression 
\cite{14}, heart rates~\cite{15}, cardiovascular system~\cite{16}, kidney structure~\cite{17}, 
cellular differentiation~\cite{18}, neuron branching~\cite{19}, and image rendering, 
image processing, mammography, images of human brains~\cite{20,21}, as
well as to the classification of strokes in brains~\cite{22}.  Since fractals in 
Nature are directly related to some growth process, therefore, methods 
such as the Multiple Reduction Copy Machine (MRCM) and the L-systems 
have also been used to generate plants, trees and bushes~\cite{4}.

Attempts have been made to introduce and use {\it practical fractals}~\cite{2},
which basically refer to fractals with a limited range of self-similarity.  On the 
other hand, multi-fractals are used to describe phenomena whose different 
components may have different scaling exponents, which may require a 
spectrum of exponents~\cite{2,3,4}.  Moreover, it was also shown that a 
different family of fractals can also be defined by removing one fractal 
property from the list of fractal basic characteristics; for example, in 
intelligent processing systems, a fractal of limited scale range and partial 
symmetry is called {\it semi-fractal}~\cite{23}.  Similar ideas were used 
in studies of images, structures and even sounds~\cite{24,25}, but to the 
best of our knowledge, they have never been applied to biology and its 
systems.

Despite a broad range of applications of fractals to biology and bio medicine, 
we propose here to refine the idea of fractals by introducing the concept of 
{\it reduced fractals} with specific limited scale range and only partial 
self-similarity (see Section~\ref{sec2}).  Moreover, we demonstrate how to 
construct spectra of these reduced fractals and apply them to determine 
different levels of complexity of natural phenomena.  Our specific application 
involves {\it green algae}~\cite{26,27}, whose scales and self-similarities show 
significant variations.  The main purpose of this paper is not to perform detailed 
studies of all known algae, but instead use selected green algae to justify the 
need for spectra of reduced fractals and demonstrate their advantage over 
the previous use of fractals in biology. We also suggest that the obtained 
spectra can be used to classify the considered algae by identifying spectra 
associated with them.

Our paper is organized as follows: in Section~\ref{sec2}, reduced fractals are defined 
and applied to selected biological systems; fractal dimensions for the considered 
biological systems and the resulting spectra of reduced fractals for these systems 
are presented in Section~\ref{sec3}; our conclusions are given in Section \ref{sec4}. 

\section{Reduced Fractals in Biology}\label{sec2}

As first pointed out by Mandelbrot~\cite{1}, most natural structures do
not show self-similarity at infinitely many stages, as classical fractals do,
but instead their self-similarity occurs only at a finite number of stages.
Moreover, there can be imperfections in self-similarity resulting from the
fact that a smaller cluster is unlikely to be exactly the same as a larger 
cluster, in other words, self-similarity is only approximate~\cite{3,4,5,6}.  
If there are variations in miniature copies, then self-similarity is 
statistical~\cite{2,3}; however, if miniature copies are distorted (skewed), 
then self-similarity becomes self-affinity~\cite{2}. This shows that in 
natural structures the range of magnification is also finite~\cite{4}.

The presence of these limitations leads us to believe that it is necessary 
to provide an integrated framework towards a definition of fractals 
applicable to natural structures, and such a framework is established in 
this paper by introducing the concepts of {\it reduced fractals} and their 
corresponding {\it spectra}, which for practical reasons must be discrete.  
The reduced fractals considered in this paper all have four basic characteristics 
described in Section~\ref{sec1}, but two of them, namely, self-similarity and range 
of magnification, are finite; this makes our definition consistent with 
practical fractals introduced earlier~\cite{2}, but different than the concept 
of semi-fractal~\cite{23}.  Our definition also allows for {\it wild fractal}
or fractals whose self-similarity is limited to 1 or 2 stages; nevertheless,
some scaling properties of these fractals can still be identified~\cite{4}.  In 
the following, we identify reduced fractals in selected biological systems.

\section{Applications to Biology}\label{sec3}
\subsection{Selected Biological Systems and Their Self-Similarity}\label{sec3.1}

Algae are very simple plants that can range from the microscopic, to
large seaweeds. It's very diverse and found everywhere, from being the 
ingredient used to thicken ice cream to producing  70\% of the air we 
breathe. This diversity is reflected in the enormous variation exhibited 
by their morphological and physiological traits.There are several methods 
for algae identification such as genetic methods. However, such approach 
require time-consuming operations and becomes impractical for large-scale 
identification in fields such as food authentication. It becomes critically 
important to identify algae without compromising food safety and to 
meet the economic demands. With 37,000 algae species, using fractals 
makes it possible.

\begin{figure}
\begin{center}
\includegraphics[width=\linewidth]{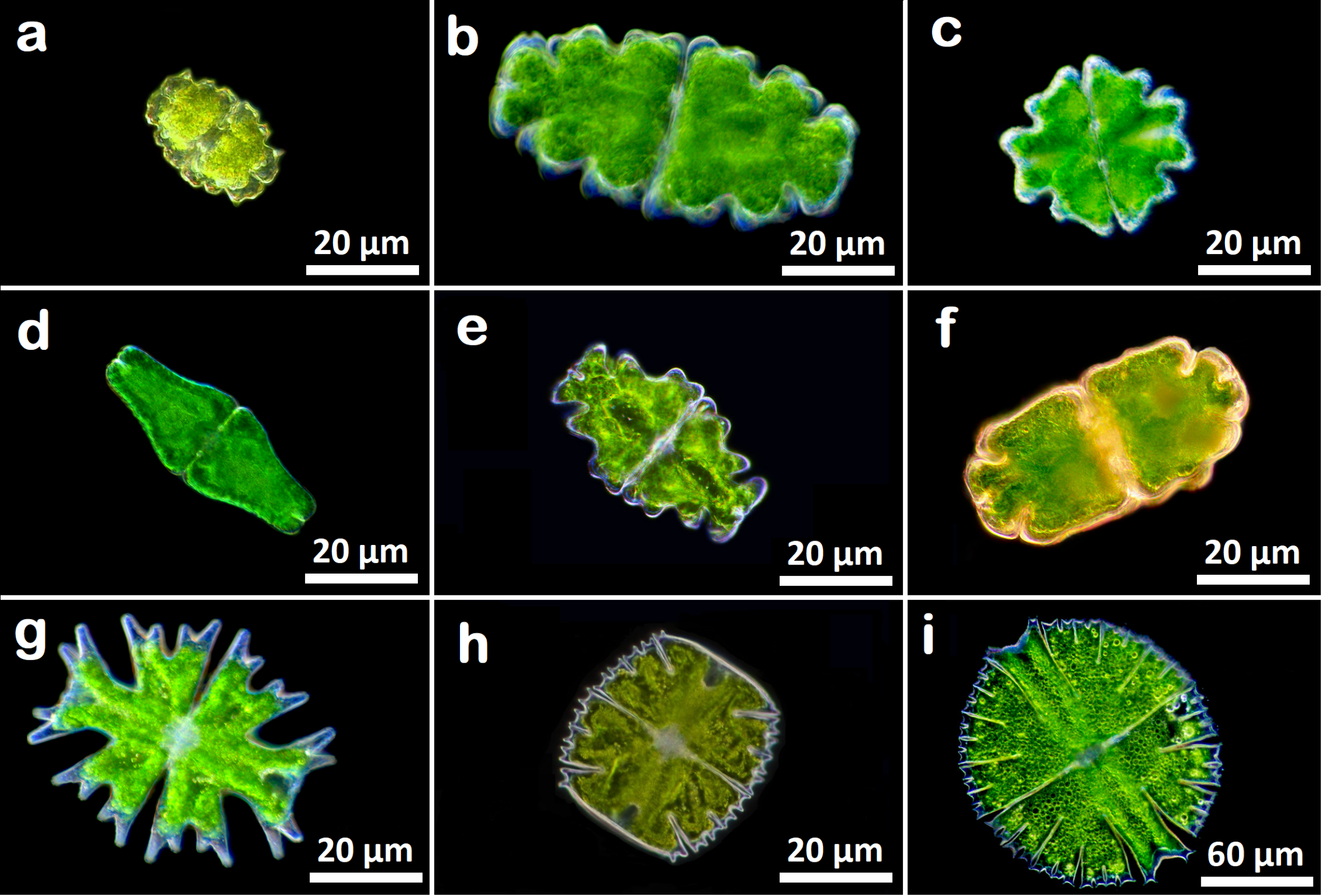}
\end{center}
\caption{
(\textbf{a}) {\it Volvox globator}   $F_d=1.2288$, 
(\textbf{b}){\it Volvox aureus}     $F_d=1.3701$,
(\textbf{c}) {\it Eudorina elegans}  $F_d=1.6975$,
(\textbf{d}) {\it Scenedesmus granlulatas} $F_d=1.7097$,
(\textbf{e}) {\it Pediastrum clothratum} $F_d=1.7182$,
(\textbf{f}) {\it Pediastrum angulosum}  $F_d=1.7806$,
(\textbf{g}) {\it Desmodesmus magnus}    $F_d=1.7447$ and
(\textbf{h}) {\it Tetraedron minimum}    $F_d=1.7087$.
}\label{fig1}
\end{figure}

\begin{figure}
\begin{center}
\includegraphics[width=\linewidth]{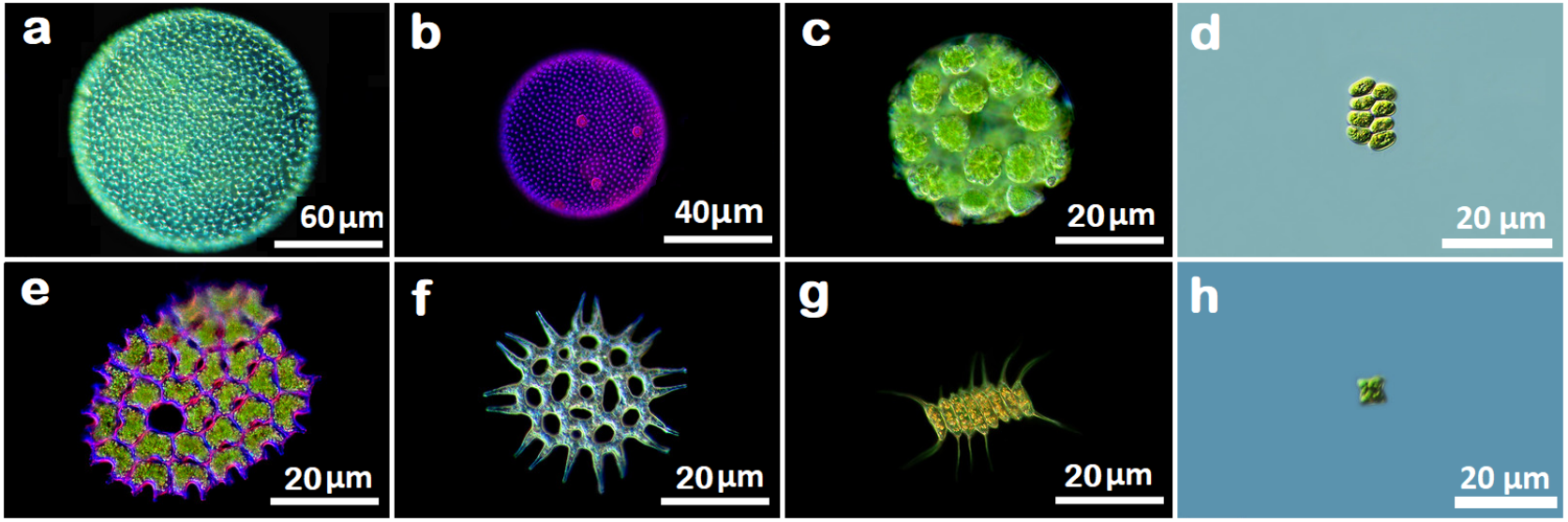}
\end{center}
\caption{Algae of Class Chlorophyceae
(\textbf{a}) {\it Volvox globator}   $F_d=1.2288$, 
(\textbf{b})~{\it Volvox aureus}     $F_d=1.3701$,
(\textbf{c}) {\it Eudorina elegans}  $F_d=1.6975$,
(\textbf{d}) {\it Scenedesmus granlulatas} $F_d=1.7097$,
(\textbf{e}) {\it Pediastrum clothratum} $F_d=1.7182$,
(\textbf{f}) {\it Pediastrum angulosum}  $F_d=1.7806$,
(\textbf{g}) {\it Desmodesmus magnus}    $F_d=1.7447$ and
(\textbf{h}) {\it Tetraedron minimum}    $F_d=1.7087$.
}
\label{fig2}
\end{figure}

As examples of biological systems considered in his paper, we select 
{\it green algae} (\mbox{Division}: Charophyta and Chlorophyta), 
and consider two classes of Charophyta, namely, Charophyceae and 
Zygnematophyceae, and three classes of Chlorophyta, namely, 
Chlorophyceae, Ulvophyceae and Trebouxiophyceae~\cite{26,27}. 
In Figure~\ref{fig1}, we present nine selected algae of the class Zygnematophyceae, 
and Figure~\ref{fig2} shows eight selected algae of Chlorophyceae; algae of other 
classes are used in the spectra of reduced fractals described in Section~\ref{sec3.3}.  
As shown by the images presented in {Figures}~\ref{fig1} and~\ref{fig2}, 
we consider only green algae that mainly inhabit in freshwater, but five 
different classes, with each class containing algae of different sizes, shapes 
and nature.  The considered sample of algae is diversified and rich in its 
structure, and thus it is sufficient to illustrate all the main concepts and objectives 
of this~paper.  

All considered algae of the class Zygnematophyceae are {\it 
unicellular} and they belong to the family Desmidiaceae; each
one of them splits into two parts that have perfect symmetry; 
however, if we scale down, then they show no self-similarity.  
Nevertheless, these algae are examples of reduced fractals defined 
in Section~\ref{sec2}.   On the other hand, two selected algae of the class 
Chlorophyceae that are {\it multicellular} show more prominent 
self-similarity than the unicellular algae, but still their self-similarity 
is limited, and their range of magnification is finite, which means 
that these algae can also be represented by reduced fractals.

The limits on self-similarity and on the range of magnification in
the class Chlorophyceae of colonial algae are presented in Table~\ref{tab1}, 
which contains algae of different families, namely, {\it Eudorina elegans}
belongs to the family Volvocaceae, the next two algae belong to 
the family Scenedesmaceae, and the remaining three algae are 
from the family Hydrodictyaceae.  For the algae in Table~\ref{tab1}, it is 
seen that the replication of colony (shown by magnification) is 
an indicator of limited self-similarity.  Since both self-similarity 
and the range of magnification are finite, these algae can also 
be described by reduced fractals.

\begin{table}
\caption{Self-similarity and magnification in some Chlorophyceae colonial algae\label{tab1}}
\centering
\begin{tabular}{ccccc} 
\hline\hline   
{\bf Genus sp.} & {\bf Colony Size} & {\bf Magnification}\\ 
\hline  
{\it Eudorina elegans}& 16,32& 2\\
{\it Desmodesmus magnus}& 4,8,16& 3\\
{\it Scenedesmus granlulatas}& 2,4,8,16,32& 5\\
{\it Pediastrum angulosum}& 4,8,32,64,128 & 5\\
{\it Pediastrum clothratum}& 8,16,32,64& 4\\
{\it Tetraedron minimum}& 2,4,8,16& 4\\ 
\hline  
\end{tabular}
\end{table}

The above description shows that the algae selected for this 
paper have self-similarities ranging from very limited, as is in 
the case of the unicellular algae of the class Zygnematophyceae,
to more moderate, as observed in the multicellular algae of the 
same class and in the colonial algae of the class Chlorophyceae; 
similar limitations and variations are observed in the range of 
magnification.  The observed self-similarity is limited, and it is 
not perfect, as there are variations in miniature copies, so 
self-similarity observed in the selected algae is limited and 
statistical.  For these reasons, the algae considered in this 
paper are well represented by reduced fractals.

\subsection{Fractal Dimension and Box-Counting Method}\label{sec3.2}

The sample of selected algae allowed us to identify reduced 
fractals as the best way to represent them and describe their 
physical properties; one such property is the irregularity, or 
complexity, of their surface and structure.  According to
Mandelbrot~\cite{1}, the complexity can be measured by 
the so-called {\it fractal dimension, FD}, which is a bounded 
set $\mathcal{S}$ in Euclidean $n$-space and is defined as 

\begin{equation}
FD = \lim_{r \rightarrow 0} \frac {\log_{10} (N_r)}{\log_{10} 
(1/r)}\ , 
\label{eq1}
\end{equation}

where $N_r$ represents a number of distinct copies of $\mathcal{S}$
in the scale $r$ ~\cite{29}.  Moreover, the union of $N_r$ copies must 
cover the set $\mathcal{S}$ completely.

The $FD$ can be calculated for deterministic fractals and if 
an object has deterministic self-similarity, its $FD$ is the same as 
its box-counting dimension $BCD$~\cite{29}.  However, biological 
systems are not ideal deterministic fractals.  Therefore, $BCD$ 
computed by the box-counting method is only an estimate of 
$FD$.  Nevertheless, the box-counting method is one of the most 
commonly used techniques to calculate $FD$ for images~\cite{29}.  
The method is also adapted in this paper to perform calculations of 
the fractal dimension for images of the selected green algae (see 
Section \ref{sec3.1}).

The images of the considered green algae are planes with 
the pixel position denoted by the coordinates $(x,y)$, and with 
the third coordinate $(z)$ denoting pixel gray level. In the 
box-counting method, the plane $(x,y)$ is partitioned into 
separate blocks of size $\lambda \times \lambda$ with 
$\lambda$ being an integer and $\lambda = r$.  As shown 
by Equation (\ref{eq1}), the box-counting method requires 
$N_r$, which is found in the following way~\cite{29}.  Boxes 
of size $\lambda \times \lambda \times \lambda^{\prime}$, 
{where} $\lambda^{\prime}$ is the height of each box 
associated with the gray level, are stuck on top of each 
other above each block. Then, the number of boxes, $n_r$, 
covering each block is given by 

\begin{equation}
n_r (i,j)= 1 - k + l\ , 
\label{eq2}
\end{equation}

where $k$ and $l$ represent the minimum and maximum 
gray levels in the $(i,j)$th block that go in the $k$th and 
$l$th boxes, respectively~\cite{29}.  Then, $N_r$ is calculated 
for different values of $r$ by taking into account the 
contributions from all the blocks  

\begin{equation}
N_r = \sum_{i,j} n_r (i,j)\ ,
\label{eq3}
\end{equation}

which allows estimating the $FD$ from the least squares 
linear fit of $\log_{10} (N_r)$ plotted versus $\log_{10}(1/r)$.  
In the specific practical implementation of the box-counting 
method in this paper, we followed~\cite{30}.

Now, in this approach, the slope of the line equals $FD$ and 
it is defined as the amount of change along the $\log_{10} (N_r)$-axis, 
divided by the amount of change along the $\log_{10}(1/r)$-axis.  
The resulting slopes and fractal dimensions range between $1$ 
and $2$ for this kind of analysis, which corresponds to the range 
between a line that is straight with its \mbox{dimension = $1$} and a 
line that is so wiggly that it completely fills up a 2-dimensional 
plane.  This means that when the slope becomes steeper, then the 
$FD$ of such an image is larger because of its higher complexity.  
On the other hand, when the slope is flatter (closer to a straight-line), 
then the fractal dimension is smaller, as it reflects the image of lower 
complexity, which implies that the amount of detail grows slowly 
with increasing magnification.

Since in this paper, we consider some algae that show very 
limited self-similarity, all results presented below are obtained 
by performing calculations of $FD$ by using the box-counting 
method.  The computed fractal dimension by this method is a 
metric that characterizes algae complexity or space-filling 
characteristic.  As already pointed out~\cite{4,28}, most previous 
studies failed to evaluate the assumption of statistical self-similarity 
that underlies the validity of the method.  Another source of error 
is arbitrary grid placement, which is strictly positive and varies as 
a function of scale, which may make the procedure's slope estimation 
step non-unique~\cite{4}.  In our calculations performed in this paper, 
both errors are eliminated by the box-counting method described above.

\subsection{Fractal Dimension for Selected Algae}\label{sec3.3}

The results presented in Table~\ref{tab2} show that the unicellular algae 
of the the class Zygnematophyceae and the family Desmidiaceae 
have high fractal dimensions ranging approximately from $1.8$ 
to $1.9$; this narrow range of the $FDs$ implies that the level of 
complexity (or irregularity) of surfaces and structures is very 
similar for all these selected objects.  The results are also 
consistent with the fact that the unicellular algae have very limited 
ranges of self-similarity and magnification that are observed in these 
objects as is already pointed out in Section \ref{sec3.1}.  It must also be 
noted that the unicellular algae in Table~\ref{tab2} are separated 
into two groups called here {\it Euastrum} and {\it Micrasterias}, 
and that within each group, algae are ordered based on their increasing 
fractal dimension.  Since all algae shown in Table~\ref{tab2} belong 
to the same family Desmidiaceae, similar ordering can be made for 
other families of green algae. 

\begin{table}[H]
\caption{Fractal dimension of unicellular algae of the class Zygnematophyceae and its family Desmidiaceae\label{tab2}}
\centering
\begin{tabular}{ccccc}
\hline\hline   
{\bf Genus species} & {\bf Cell shape} & {\bf Fractal dimension}\\ 
\hline
{\it Euastrum oblongum}	   & Ellipsoid& $1.8598$\\
{\it Euastrum verrucosum}   & Ellipsoid& $1.8739$\\
{\it Euastrum ansatum}	   & Ellipsoid& $1.8801$\\
{\it Euastrum humerosum}   & Ellipsoid& $1.8907$\\
{\it Euastrum crissum}	   & Ellipsoid& $1.8948$\\ 
{\it Micrasterias americana} & Spherical& $1.8117$\\
{\it Micrasterias truncata}    & Spherical& $1.8703$\\
{\it Micrasterias rotata}	   & Spherical& $1.8749$\\ 
\hline
\end{tabular}
\end{table}

The fractal dimension in Table~\ref{tab3} is computed for multicellular 
and colonial algae of three families (Volvocaceae, Scenedesmaceae 
and Hydrodictyaceae) of the class Chlorophyceae, and the 
results are presented for the selected members of each class, 
which are ordered based on their increasing fractal dimension.  
The presented results show that the $FD$ of the two
multicellular algae ({\it Volvox globator} and {\it Volvox aureus}) 
are the lowest, which means that the complexity of these two algae 
is the highest among the selected objects.  Interestingly, the 
$FD$ of {\it Eudorina elegans} is more similar to one member 
of the family Scenedesmaceae and one member of the family 
Hydrodictyaceae, which is caused by the fact that these algae 
are colonial despite being members of different families.

Similarly, fractal dimension of {\it Tetraedron minimum}, which 
belongs to the family Hydrodictyaceae, is closer to that of the 
members of the family Scenedesmaceae.  The reason is likely 
caused by the fact that these algae are colonial, but also by 
similarities in their cell shapes.  However, the two remaining 
members of the family Hydrodictyaceae have significantly 
higher fractal dimensions that may be caused by both different 
cell and colony shapes between {\it Tetraedron minimum} 
and these two members.

\begin{table}
\caption{Fractal dimension of colonial and multicellular algae of the class Chlorophyceae 
and its three different families.\label{tab3}}
\centering
\begin{tabular}{ccccc} 
\hline\hline   
{\bf Family} & {\bf Genus sp.} & {\bf Form} & {\bf Cell shape}& {\bf Fractal dimension}\\ 
\hline  
Volvocaceae 	&{\it Volvox globator}			&Multicellular& Spherical & 1.2288\\
Volvocaceae 	&{\it Volvox aureus}			&Multicellular& Spherical & 1.3701\\ 
Volvocaceae       &{\it Eudorina elegans}		&Colonial	& Spherical & 1.6975\\
Scenedesmaceae &{\it Scenedesmus granulatus}	&Colonial	& Ellipsoid & 1.7097\\
&               	 &                			   			&Crescent	&\\
Scenedesmaceae &{\it Desmodesmus magnus}	&Colonial	& Ellipsoid & 1.7447\\
Hydrodictyaceae&{\it Tetraedron minimum}     	&Colonial	& Ellipsoid & 1.7087\\
&                	&      				         		&Spherical	&\\
Hydrodictyaceae&{\it Pediastrum clothratum}	&Colonial	& Oval & 1.7182\\
Hydrodictyaceae&{\it Pediastrum angulosum}	&Colonial	& Cylindrical & 1.7806\\
\hline  
\end{tabular} 
\end{table}

The results presented in Tables~\ref{tab2} and~\ref{tab3} 
demonstrate that algae can be ordered within each family by 
using their $FD$.  This may be useful in classification of algae 
and their studies, since typically algae within a given family 
are neither ordered nor organized~\cite{29}. The proposed order of 
decreasing complexity can be replaced by increasing complexity, which 
would require us to reverse the orders in 
Tables~\ref{tab2} and~\ref{tab3}.

The validity of the computed fractal dimensions must be verified 
by comparing our results to those obtained before, specifically, for 
different algae.  To the best of our knowledge, no FDs were calculated 
for the set of algae selected for this paper.  However, for {\it 
Cladophora rupestris} of the class Chlorophyceae, $FD = 1.76$ was 
obtained~\cite{31}, which is consistent with the values of Table \ref{tab3} 
for the colonial forms.  Independent calculations of FD for {\it Cladophora rupestris} 
done in~\cite{32} gave $FD = 1.59$, which slightly differs from the results obtained 
in~\cite{31} and in Table \ref{tab3}; the main reasons for the difference are 
improvements in modern computations as compared to those preformed almost 
25 years ago~\cite{32}.

The calculations of FDs were also done for {\it Laminaria digitata} and 
{\it Fucus serratus} of the class Phaeophyceae, for which $FD = 1.23$ and 
$1.11$ were obtained, respectively~\cite{31}.  These values are close to the 
FDs of Table \ref{tab3} found by us for the multicellular forms. Moreover, the 
computations of FDs for 16 selected brown algae were also done~\cite{33}, 
and the obtained results range between $FD = 1.3$ and $FD = 1.7$, which 
is consistent with the results of Table~\ref{tab3}. Thus, there is an agreement 
between the previously obtained results~\cite{31,32,33} and the results presented 
in this paper.  However, it must be pointed out that direct comparisons 
cannot be done because there are no computations of FDs for green algae
selected by us for our investigation.

\subsection{Spectra of Reduced Fractals for Selected Green Algae}\label{sec3.4}

\begin{figure}[H]
\includegraphics[width=\linewidth]{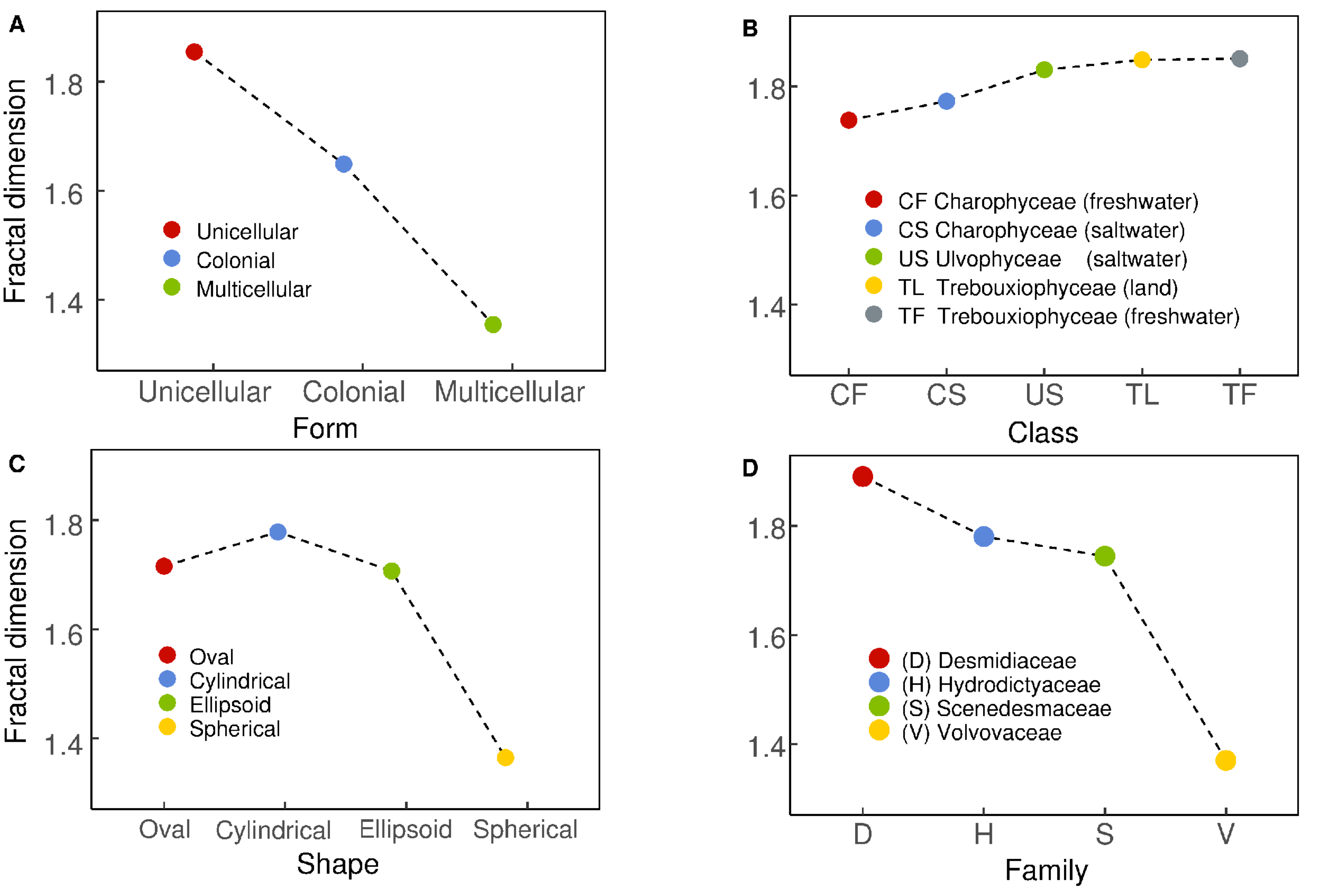}
\caption{Spectra of reduced fractals for algae of different form (panel (\textbf{A})), 
class (panel (\textbf{B})), shape (panel (\textbf{C})), and family (panel (\textbf{D})).}
\label{fig3}
\end{figure}

The fractal dimensions calculated for different green algae can now 
be used to obtain spectra of reduced fractals (SRFs).  The spectra 
are generated by plotting the fractal dimensions versus selected 
characteristics of algae.  Among different characteristics, we 
consider forms and shapes of algae described in Section~\ref{sec3.1}.  
Moreover, we also demonstrate how to generate SRFs for algae 
of different classes and families.  In the panels A, B, C and D of 
Figure~\ref{fig3}, we show the SRFs for algae of different forms, classes, 
shapes and families.  Let us now describe each panel of Figure~\ref{fig3} 
and discuss the biological implications of the presented~SRFs.

Panel A of Figure~\ref{fig3} shows the form of the SRF for one 
unicellular alga of the class Zygnematophyceae and the family Desmidiaceae, 
and one colonial and one multicellular algae of the class Chlorophyceae,
whose families are Scenedesmaceae and Volvocaceae, respectively.
As expected, the presented SRF demonstrates a rapid decrease of 
fractal dimension with increasing algae's complexity.  The observed
almost linear spectrum allows us to establish the following criteria 
for classifying green algae: $FD > 1.8$ for unicellular, $FD > 1.6$ 
for colonial, and $FD < 1.4$ for multicellular.

The SRF presented in panel B of Figure~\ref{fig3} shows high fractal 
dimensions $FD > 1.7$ for all considered algae, which are two algae of 
the class Charophyceae, one alga of the class Ulvophyceae, and two algae 
of the class Trebouxiophyceae.  The resulting spectrum shows that differences 
between freshwater and saltwater (or land) within the same class are 
practically negligible, which is shown by flat parts of the SRF.  Moreover, 
the entire spectrum remains practically flat, which implies that differences 
in the fractal dimension (or complexity) between the considered three 
classes are small.  

The fact that algae have different shapes is well-known \cite{26,27}.
Using several most commonly known shapes of green algae (see 
also Figures~\ref{fig1} and~\ref{fig2}, and Table~\ref{tab3}), 
we obtained the RFS shown in panel C of Figure~\ref{fig3}.  
The spectrum shows its maximum for cylindrical algae, and smaller 
values of the fractal dimension for oval and ellipsoid algae; both 
oval and ellipsoid algae have comparable fractal dimensions.  An 
interesting result is that spherical algae have the lowest fractal 
dimensions, which implies that their complexity is the highest 
among algae considered for this spectrum.   

The panel D of  Figure~\ref{fig3} shows the SRF for different families.  
The presented results are consistent with the SRFs given in the previous 
panels, specifically, in panel A, as the SRF of panel D mainly 
reflects differences between algae being unicellular (Desmidiaceae), 
colonial (Hydrodictyaceae and Scenedesmaceae) or multicellular 
(Volvocaceae).  The spectrum demonstrates (similar to the SRF of panel A) 
a rapid decrease in the fractal dimension, and related 
increase in algae complexity, for colonial algae as compared to 
unicellular ones.  An even sharper decrease in the fractal dimension 
is observed for multicellular algae as compared to colonial.  As 
expected, there is only a small difference in the fractal dimension 
between the families Hydrodictyaceae and Scenedesmaceae as 
members of these families are mainly colonial (see Table~\ref{tab3}).    

\subsection{Discussion of the Obtained Results}\label{sec3.5}

The main results of this paper are fractal dimensions (FDs) given in 
\mbox{Tables \ref{tab1} and \ref{tab2}}, and the spectra of reduced 
fractals (SRFs) presented in Figure~\ref{fig3}.  The results were obtained 
for the selected green algae shown in  Figures~\ref{fig1} and~\ref{fig2}, 
and our computations were performed using the box-counting method that 
is described in Section~\ref{sec3.2}. A similar method was used in some 
previous studies to compute fractal dimensions for known images of 
different biological systems to determine their complexity, structure,
function and \mbox{organization~\cite{34,35,36}}, as well as for medical 
images \cite{37}.This common use of the fractal dimension in the work cited 
above as well as in this work has been motivated by the fact that the FD captures 
and describes the complexity of an object by providing one unique number that 
corresponds to this object, and its value determines the change in 
complexity in detail with the change in~scale.

We have also calculated the SRFs and demonstrated that they can be
used as a new tool to investigate properties of green algae and also to 
classify them, based on the form of their SRF, within their families as 
currently algae of a given family are neither ordered nor organized \cite{26,27}.  
Moreover, the SRFs can also be generated for other types of algae, namely, 
Macroalgae (red and brown) or Microalgae \cite{26,27}, as well as for many 
other diversified biological systems, such as the roots of plants \cite{28}
and their complexity \cite{34}, scaling time in biochemical networks \cite{35}, 
organization of ecosystems \cite{36}, human physiology and well-being \cite{37},
and microbial colonies \cite{38}. We do hope that biologists working in 
different areas, and other natural scientists, find the SRFs useful 
in their work and apply them to different natural systems.

The main advantage of using SRFs is that the shapes of these 
spectra change from one biological system to another, which makes 
it easy to identify different systems by the characteristic shapes 
of their spectra. As a result, it is suggested that the spectra may 
be used as a tool to classify different systems, and also to make 
comparisons between different biological systems. In other words, 
the spectra uniquely show differences and similarities between 
diverse systems, which was not the case in the previous studies that 
were limited to one particular object of a certain class or family, 
and a certain shape or form.  It must also be pointed out that the 
SRFs are an efficient and low-cost tool compared to other more 
advanced techniques, like machine learning or detailed digital 
analysis of images.

Finally, our suggestion that the SRFs can be used to classify 
different biological systems requires more studies. Specifically,
a test analysis and a confusion matrix analysis are needed to 
formally demonstrate the validity of our suggestion; however, these
topics are of the scope of this paper and they will be considered
elsewhere.

\section{Conclusions}\label{sec4}

The concept of reduced fractals, with a specific limited scale range and only partial 
self-similarity, is introduced and used to generate spectra of reduced fractals.  
To demonstrate the applicability of these spectra to biology, the spectra are 
generated for selected green algae, which include Charophyta and Chlorophyta
algae, and their classes Charophyceae, Zygnematophyceae, Chlorophyceae, 
Ulvophyceae and Trebouxiophyceae.  By showing how these spectra can be used
to investigate physical properties of algae and to classify them within their families,
we hope that the spectra will become a new tool to study algae, including also 
red and brown algae, as well as microalgae.  It is also suggested that the spectra 
can be used for other biological systems, whose images are known, and that they 
may provide biologists with a tool to bridge over to physics, electro-sensory artificial 
life and synthetic biology. The spectra may also become a useful tool in other natural 
sciences.

\end{document}